\documentclass[10pt]{article}
\usepackage[cp1251]{inputenc}
\usepackage[english]{babel}
\usepackage{graphicx}
\usepackage{amssymb}
\usepackage{amsmath}
\usepackage{amsthm}
\usepackage{amsfonts}
\usepackage{amssymb}
\title{{\bf \Large {\centering Cosmological Model of Interacting Tachyon Field }}\\ \vspace{5mm}
{\normalsize {\bf V.\,K. Shchigolev}\thanks{E-mail:
vkshch@yahoo.com},~~{\bf M.\,P. Rotova
\thanks{E-mail: canopus-007@yandex.ru}}}\\\vspace{5mm} {\small ~~Department of
Theoretical Physics, Ulyanovsk
State University,\\42 L. Tolstoy Str.,
Ulyanovsk 432000, Russia\\
\bigskip
\begin{quote} {\bf Abstract} -- In this paper we investigate a tachyon field model in cosmology, provided its interaction with the quintessence or phantom fields.The model  takes into account this interaction beyond the usual  approach, in which the interaction is phenomenologically  described by the energy flow between the matter components. In our model, the interaction of tachyon field with a canonical scalar field is taken into account through the interaction potential in the total Lagrangian of the system, like in the case of two or more canonical scalar fields. We obtain the different types of exact solution for the  model by employing the so-called "first order formalism" procedures.\\
\bigskip
\flushleft {{\bf PACS}: 98.80.-k,98.80.Jk.}\\
{\bf Key words}: Cosmological model, tachyon field, interaction, exact solutions.\\
\end{quote}
}}
\date{}
\begin{document}
\maketitle
\vspace{-2cm}
\section{Introduction}

\quad\, Several cosmological observations, such as SNe type
Ia \cite{C1,C2}, Cosmic Microwave Background Radiation \cite{C3,C4}, Baryon
Acoustic Oscillations in galaxy surveys \cite{C5,C6} etc., indicate that the observable Universe is undergoing a phase of accelerated expansion. In order to explain such phenomena, one needs to add an extra energy component of the Universe, which has a large negative pressure and also dominates over all other components e.g. non-relativistic matter, radiation etc. As known, this mysterious component is usually referred to as "dark energy" (DE).

To accelerate the Universe expansion, equation of state (EoS) of DE,  $w=p/\rho$ where $p$ and $\rho$ are the pressure and energy density respectively, must satisfy $w <-1 / 3$. The simplest candidate for the role of DE is the cosmological constant,
for which EoS is $w = -1$. However, there is some evidence that DE may evolve from $w> -1$ in the remote past to $w <-1$ at present. To date, there have been studied a large class of scalar - field models of DE \cite{C7},  including  tachyon \cite{C8}, phantom \cite{C9}, quintom \cite{C10} and other models. Furthermore, the proposals for DE include interacting DE model \cite{C11},  braneworld models \cite{C12} and holographic DE models\cite{C13}. Phantom field (with negative kinetic energy) was also supposed as a candidate for DE, since it allows sufficient negative pressure with EoS $w <-1$ \cite{C14,C15}. The so-called quintom scenario of DE has been developed in order to explain the dark energy EoS crossing $-1$. The first
model of quintom scenario of dark energy has been obtained in \cite{C10}.
This model is studied in detail later e.g in \cite{C16,C17}.

Recently there has been increasing interest for constructing  the tachyon models in
cosmology \cite{C18}, where the appearance of a tachyon field is mainly motivated by the string theory \cite{C19}.  In the context of string theory,
the tachyon field in the world volume theory of the open string stretched between a D-brane
and an anti-D-brane or a non-BPS D-brane plays the role of scalar field in the quintom
model. The effective action used in the study of tachyon cosmology consists of the
standard Einstein-Hilbert action and an effective action for the tachyon field on unstable
D-brane or D-brane anti D-brane system. What distinguishes the tachyon action from the
standard Klein- Gordon form for scalar field is that the tachyon action is of the " Dirac-Born-Infeld " (DBI) form. As known, tachyon fields are able to contribute for the inflation and late time acceleration \cite{C20}. The tachyon model of inflation has been discussed in \cite{C21,C22}. Tachyon dark energy as a function of various forms of its potential was studied in \cite{C23}-\cite{C26}. A model for tachyon field with non-minimal derivative  coupling to gravity has been considered in \cite{C27}.

In view of the latest developments in DE modeling, the study of interacting tachyon  models acquired much significance. It seems natural to study different hybrid models which could include an interacting tachyon field in the framework of different approaches.  However, the model accounting for the interaction of tachyon field
with other fields has a well-known problem. So far the description  of tachyonic interaction is made phenomenologically by involving an energy exchange term to govern the interaction (see, e.g., \cite{C28}-\cite{C44}).

In this paper we made an attempt to develop an alternative approach for description of the interacting tachyon field, which is based on a more conventional phenomenological description via an interaction potential. We obtain the different types of exact solution for the Friedmann-Robertson-Walker (FRW)  model filled with the interacting  tachyon and quintessence (phantom) scalar fields by employing the so-called first order formalism  procedure \cite{C45,C46}.

\section{Basic equations}
\quad
An effective DBI Lagrangian density inspired from the super-string theory
is
\begin{equation}\label{1}
{\cal L}_{DBI} = -f^{-1}(\phi)\sqrt{\displaystyle 1-f(\phi)(\nabla \phi )^{2}}+f^{-1}(\phi)-U_1(\phi),
\end{equation}
where the tension of the brane is $f^{-1}(\phi)=T_3 h^4(\phi)$ with $h(\phi)$ being the warped factor in the the metric
\begin{equation}\label{2}
ds^2_{10}=h^2(r)ds^2_4+h^{-2}(r)(dr^2+r^2ds^2_5),
\end{equation}
and $U_1(\phi)$ is the potential for the DBI-field, arising from quantum interactions between the D3-brane associated with $\phi$, and other D-branes. In terms of the brane tension $T(\phi)$, the Lagrangian density is
\begin{equation}\label{3}
{\cal L}_{\phi} = -T(\phi)\sqrt{1-\dot \phi ^{2}/T(\phi)}+T(\phi)-U_1(\phi),
\end{equation}
for a homogeneous and isotropic Universe. By adding to the tachyon component a scalar field $\varphi$ described by the Lagrangian density with an interaction potential
\begin{equation}\label{4}
{\cal L}_{\varphi}=\epsilon \frac{1}{2}\dot \varphi^2-U_2(\varphi)-U_{int}(\phi,\varphi),
\end{equation}
we have the total Lagrangian density ${\cal L}={\cal L_{\phi}}+{\cal L_{\varphi}}$ for the hybrid source of gravity. Here,  $ \epsilon = + 1$  represents  quintessence while  $ \epsilon = - 1$ refers to phantom field.

Relying on a phenomenological approach in the presence of scalar field, we assume further that the brane tension depends also on $ \varphi $ as a parameter. If we substitute $T(\phi, \varphi)= U_1(\phi) + U_2(\varphi)+U_{int}(\phi,\varphi)$ into the total Lagrangian density and denote $T(\phi, \varphi)=V(\phi, \varphi)$, i.e.
\begin{equation}\label{5}
V(\phi, \varphi)=U_1(\phi) + U_2(\varphi)+U_{int}(\phi,\varphi),
\end{equation}
then we can write down the following Lagrangian density for our model:
\begin{equation} \label{6}
{\cal L}=-V(\phi,\varphi )\sqrt{1-\frac{\dot{\phi }^{2} }{V(\phi ,\varphi )} } +\frac{\epsilon }{2} \dot{\varphi }^{2}.
\end{equation}
Some motivation for this model is that in the case of slow roll, viz.  $\dot{\phi }^{2}<<V(\phi ,\varphi )$, the total Lagrangian is reduced to the canonical quintom model.

Keeping in mind that all quantities, $\phi(t), \varphi(t)$ and $V(\phi(t),\varphi(t))$ are functions of time $t$ which should be found from the set of field equations, we can suppose the re-definition of tachyon field as follows (cf \cite{C47}):
\begin{equation}\label{7}
\dot\chi^2(t)=\frac{\dot\phi^2(t)}{V(\phi(t),\varphi(t))} \Leftrightarrow    \phi \to \chi  =\pm \int \frac {\dot\phi(t)\, d\,t}{\sqrt{V(\phi(t),\varphi(t))}}.
\end{equation}
Substitution of the latter into (\ref{6})  leads to the following Lagrangian density:
\begin{equation} \label{8}
{\cal L}=-V(\chi,\varphi )\sqrt{1-\dot{\chi }^{2}} +\frac{\epsilon }{2} \dot{\varphi }^{2}.
\end{equation}
For the sake of simplicity, we
consider a spatially flat FRW cosmological model with the
space-time interval
\begin{equation}
\label {9} d s^2 = d t^2- a^2 (t)\left(d r^2+r^2d \Omega
^2\right),
\end{equation}
where $a(t)$ is a scale factor.  Taking into account this metric, we can derive from Lagrangian (\ref{8}) the following equations for the tachyon and scalar fields:
\begin{equation}\label{10}
\frac{\ddot{\chi }}{1-\dot \chi^2}+3H\dot{\chi} + \frac{V_{\chi}}{V}+\dot \chi \, \dot \varphi \frac{V_{\varphi}}{V}=0,
\end{equation}
\begin{equation} \label{11}
\ddot{\varphi }+3H\dot{\varphi }+\epsilon V_{\varphi}\sqrt{\displaystyle 1-\dot\chi^2}=0,
\end{equation}
where $H=\dot a(t)/a(t)$ is the Hubble function, $V_{\chi}=\partial V(\chi,\varphi)/\partial \chi $, and $V_{\varphi}=\partial V(\chi,\varphi)/\partial \varphi$.
Besides, with the help of well-known formula for the energy-momentum tensor,
$T_{\mu \nu } =2\partial {\cal L}/\partial g^{\mu \nu } -g_{\mu \nu } {\cal L}$,
and from (\ref{8}) it follows that the total effective energy density and pressure are
\begin{equation} \label{12} \rho _{tot} =\frac{V(\chi,\varphi)}{\sqrt{1-\dot{\chi }^{2}} } +\frac{\epsilon }{2} \dot{\varphi }^{2} ,
 \,\,\, p_{tot} =-V(\chi,\varphi )\sqrt{\displaystyle 1-\dot{\chi }^{2} } +\frac{\epsilon }{2} \dot{\varphi }^{2}.
\end{equation}
Hence, the behavior of this model is defined by equations (\ref{10}), (\ref{11}) and the Friedmann equations, $(3/2)H^2=\rho_{tot}$ and $\dot H = -(\rho_{tot}+ p_{tot})$ where $4\pi G = 1$.  From the metric (\ref{9}) and equations (\ref{12}), one can conclude that
\begin{equation}
\label{13}
\frac{3}{2} \, H^{2} =\frac{V(\chi,\varphi)}{\sqrt{\displaystyle 1-\dot \chi^2 }} +\frac{\epsilon }{2} \dot{\varphi }^{2},
\end{equation}
\begin{equation} \label{14}  \dot H =-\frac{V(\chi,\varphi) \dot \chi^2}{\sqrt{\displaystyle 1-\dot \chi^2}} -\epsilon \dot{\varphi }^{2} .
\end{equation}
At the same time, the energy-momentum conservation law,
$$\dot \rho_{tot}+3 H (\rho_{tot}+ p_{tot})=0,$$
leads to the following equation
\begin{equation}\label{15}
\frac{\dot\chi V}{\sqrt{\displaystyle 1-\dot \chi^2 }}\Big[\frac{\ddot{\chi }}{1-\dot \chi^2}+3H\dot{\chi} + \frac{V_{\chi}}{V}\Big] + \dot\varphi\Big[\epsilon(\ddot{\varphi }+3H\dot{\varphi })+\frac{V_{\varphi}}{\sqrt{\displaystyle 1-\dot\chi^2}}\Big]=0,
\end{equation}
which is satisfied due to the field equations (\ref{10}), (\ref{11}).

\section{Exact solutions}
\subsection{The method}

\quad\, Even for the given potential  $V(\chi,\varphi)$, it is difficult to find exact solutions for our model. However, a class of exact solutions can be obtained in terms of the so-called superpotential in the first order formalism. This procedure was first performed with a single scalar field in \cite{C45}, and it was later re-opened and extended on the case of two or more fields with standard dynamics in \cite{C46}. The method of  superpotential  can be also productively applied to the quintom models (see, e.g., \cite{C48} and \cite{C49}).

It should be noted that the general first-order equations for multiple fields, that are covariant in field space, were written down in \cite{C50}.  As seems,  there exist some examples for which the first-order framework, as  written  in terms of a superpotential, fails \cite{C51}.  Nevertheless, we are able to give several examples for which the method works.

Here, we introduce the superpotential function $W(\chi,\varphi)$ by the equality
\begin{equation}\label{16}
H=W(\chi,\varphi),
\end{equation}
in which the Hubble parameter $H(t)$, as a function of time, is presumably expressed in terms of fields $\chi(t),\varphi(t)$. By inserting (\ref{16}) into (\ref{14}), one can obtain two first-order equations as follows:
\begin{equation}\label{17}
\dot \chi =-\frac{2W_{\chi}}{3W^2-\epsilon W_{\varphi}^2},\,\,\dot \varphi = -\epsilon W_{\varphi},
\end{equation}
where $W_{\chi}=\partial W/ \partial \chi$, $W_{\varphi}=\partial W/ \partial \varphi$.
The potential $V(\chi,\varphi)$ is followed from (\ref{12}), (\ref{16}) and (\ref{17}) in the form:
\begin{equation}\label{18}
V=\frac{1}{2}\sqrt{\displaystyle \Big(3 W^2-\epsilon W_{\varphi}^2\Big)^2 - 4 W_{\chi}^2},
\end{equation}

The evolutionary equation for the superpotential $W(\chi(t),\varphi(t))$ is followed from $\dot W = W_{\chi} \dot \chi + W_{\varphi} \dot \varphi$ and Eq. (\ref{17}) as
\begin{equation}\label{19}
\dot W = -\frac{2W^2_{\chi}}{3W^2-\epsilon W_{\varphi}^2} - \epsilon W^2_{\varphi}.
\end{equation}
With the help of Eqs. (\ref{7}), (\ref{17}) and (\ref{18}), it is easy to find that the original tachyon field $\phi(t)$ can be obtained from the following equation
\begin{equation}\label{20}
\dot \phi^2 = 2 W^2_{\chi} \sqrt{1-\frac{4W^2_{\chi}}{(3W^2-\epsilon W_{\varphi}^2)^2} }= 2 W^2_{\chi} \sqrt{1-\dot \chi^2}.
\end{equation}

  To analyze this model, it is useful to consider the EoS parameter  $w=\displaystyle \frac{p_{tot} }{\rho_{tot}}=-1-\frac{2}{3}\frac{\dot H}{H^2}$ and the deceleration parameter
$q=\displaystyle  - \frac{a \ddot a}{\dot a^2}=-1-\frac{\dot H}{H^2}$ followed from Eqs. (\ref{16}), (\ref{19}):
\begin{equation}\label{21}
\Big\{\begin{array}{c}
w\\q\end{array}\Big\}=-1 +\Big\{\begin{array}{c}
2/3\\ 1\end{array}\Big\}\times\left[\frac{2W_{\chi}^2}{ W^2\Big(3 W^2-\epsilon W_{\varphi}^2\Big)} +\epsilon\frac{ W_{\varphi}^2}{ W^2}\right].
\end{equation}

So we have a wide range of possibilities to solve the model equations assuming some certain dependence $W(\chi,\varphi)$. Instead, we can provide several classes of solution for the model evolving from some conditions on superpotential. Below, we show how it can be realized with the help of some ansatz for the superpotential.

\subsection{The first ansatz}

The simplest condition on superpotential may be represented as
\begin{equation}\label{22}
W_{\chi}=\lambda_1 W^m(\chi,\varphi),\,\,\,W_{\varphi}=\lambda_2 W^m(\chi,\varphi),
\end{equation}
where $m,\,\lambda_1,\,\lambda_2$ are constants. Then from (\ref{17}) and (\ref{22}), we have
\begin{equation}\label{23}
\dot \chi =-\frac{2 \lambda_1 W^{m-2}}{3-\epsilon \lambda_2^2 W^{2m-2}},\,\,\,\, \dot \varphi =-\epsilon \lambda_2 W^m.
\end{equation}
In view of Eqs. (\ref{19}) and  (\ref{22}), we can obtain the evolutionary equation for the superpotential or, keeping in mind  Eq. (\ref{16}), the Hubble parameter as follows
\begin{equation}\label{24}
\dot W = -\frac{2\lambda_1^2\, W^{2m-2}}{3-\epsilon \lambda_2^2\, W^{2m-2}}-\epsilon \lambda_2^2\, W^{2m},\,\,\,H=W(t).
\end{equation}
Integrating this equation for a given $m$, we can find the fields from Eq. (\ref{23}). Then it is possible to find the field potential (\ref{18}), the Eos index and the deceleration parameter (\ref{21}).
Let us illustrate this procedure for two different values of $m$.

\subsubsection{The case $m=0$}

In this case, it follows from Eq. (\ref{22}) that $W_{\chi}=\lambda_1,W_{\varphi}=\lambda_2$. Therefore,
\begin{equation}\label{25}
W(\chi,\varphi)=\lambda_1\chi+\lambda_2 \varphi,
\end{equation}
and the field equations (\ref{23}) become as follows
\begin{equation}\label{26}
\dot \chi =-\frac{2 \lambda_1}{3 W^2-\epsilon \lambda_2^2},\,\,\,\dot \varphi=-\epsilon \lambda_2.
\end{equation}
From Eq. (\ref{24}), one can find the following equation for the Hubble parameter (or superpotential):
\begin{equation}\label{27}
\dot H = -\frac{2 \lambda_1^2}{3 H^2-\epsilon \lambda_2^2}-\epsilon \lambda_2^2,
\end{equation}
which can be integrated in an implicit form as
\begin{equation}\label{28}
t=\frac{1}{\lambda_2^2} H+\frac{2\lambda_1^2}{\lambda_2^3\sqrt{3\lambda_2^4-6\lambda_1^2}}\Big\{\begin{array}{c}
\tan^{-1}\\ \tanh^{-1}\end{array}\Big\}\Big(\frac{3\lambda_2 H}{\sqrt{3\lambda_2^4-6\lambda_1^2}}\Big),
\end{equation}
for $\epsilon = +1$ and $\epsilon = -1$, consequently. Besides, we can obtain that
\begin{equation}\label{29}
\Big\{\begin{array}{c}
w\\q\end{array}\Big\} = -1+\Big\{\begin{array}{c}
2/3\\ 1\end{array}\Big\}\times\left[\frac{2 \lambda_1^2}{ H^2(3 H^2-\epsilon \lambda_2^2)}+\epsilon \frac{\lambda_2^2}{H^2}\right].
\end{equation}

\subsubsection{The case $m=1$}

From Eq. (\ref{22}) with $m=1$ and $d W=W_{\chi} d \chi+W_{\varphi} d \varphi$ , it follows that
\begin{equation}\label{30}
W(\chi,\varphi)=W_0 \exp(\lambda_1 \chi + \lambda_2 \varphi),
\end{equation}
Then from (\ref{23}), we have
\begin{equation}\label{31}
\dot \chi =-\frac{2 \lambda_1}{3-\epsilon \lambda_2^2} W^{-1},\,\,\dot \varphi = -\epsilon \lambda_2 W.
\end{equation}
With the help of Eqs. (\ref{18}) and (\ref{22}), we can find the following expression for the potential:
\begin{equation}\label{32}
V=\frac{W}{2}\sqrt{\displaystyle (3-\epsilon \lambda_2^2)^2 W^2-4\lambda_1^2},
\end{equation}
where $W$ is presented by (\ref{30}).
Besides, we are able to obtain the equation for the superpotential followed from (\ref{24}):
\begin{equation}\label{33}
\dot W = -\Big(\epsilon \lambda_2^2 W^2 +\frac{2\lambda_1^2}{3-\epsilon \lambda_2^2}\Big) .
\end{equation}
Integrating Eq. (\ref{33}) with $\epsilon=-1$, we have
\begin{equation}\label{34}
H=H_0\coth \Big[\lambda_2^2 H_0 (t_f-t)\Big],
\end{equation}
where $\displaystyle H_0=\frac{\lambda_1}{\lambda_2}\sqrt{\frac{2}{3+\lambda_2^2}}$, and $t_f>0$ is a constant of integration. Therefore, the scale factor can be written down as
\begin{equation}\label{35}
a(t)=a_0 \Big\{ \sinh\Big[\lambda_2^2 H_0 (t_f-t)\Big]\Big\}^{-\displaystyle\frac{1}{\lambda_2^2}}.
\end{equation}
From Eqs. (\ref{21}) and (\ref{34}), we can find that
\begin{equation}\label{36}
\Big\{\begin{array}{c}
w\\q\end{array}\Big\}=-1-\Big\{\begin{array}{c}
2/3\\ 1\end{array}\Big\}\times\lambda_2^2\Big\{\cosh\Big[\lambda_2^2 H_0 (t_f-t)\Big]\Big\}^{-2},
\end{equation}
In view of Eqs. (\ref{31}) and (\ref{34}), it can be found that
\begin{eqnarray}
 \lambda_1\chi = \ln\cosh\Big[\lambda_2^2 H_0 (t_f-t)\Big]-C, \nonumber\\
 \lambda_2\varphi = \ln\sinh\Big[\lambda_2^2 H_0 (t_f-t)\Big]+C, \label{37}
 \end{eqnarray}
where $C$ is a constant. The behavior of this model in time is plotted in Fig. 1.

\begin{figure}[t]
\centering
\includegraphics[width=90mm,height=7cm]{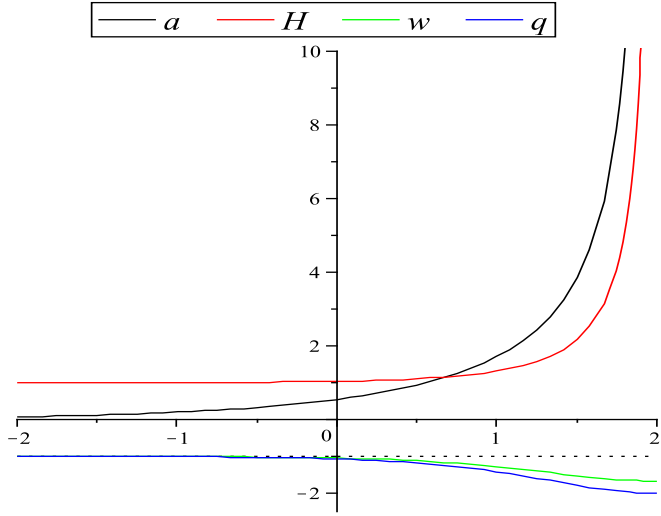}\\
\caption{The plot of the scale factor $a$, the Hubble parameter $H$, the EoS parameter $w$, and the deceleration parameter $q$ versus time $t$ for $m=1,\,\epsilon=-1$, $\lambda_1=\sqrt{2},\lambda_2=1,\, t_f=2$, and $a_0=2$.}
\label{pic1}
\end{figure}

\subsection{The second ansatz}

Let us now consider the case when
\begin{equation}\label{38}
W(\chi,\varphi)=X(\chi)\,\Phi(\varphi).
\end{equation}
In view of this ansatz and equation (\ref{17}), we have
\begin{eqnarray}
W_{\chi}=X'\Phi,\,\,\,W_{\varphi}=X\Phi',\nonumber\\
\dot \chi =-\frac{2 X'}{X^2}\,\frac{\Phi}{3\Phi^2-\epsilon \Phi'^2},\,\,\dot \varphi=-\epsilon X \Phi', \label{39}
\end{eqnarray}
where the primes denote derivatives with respect to the appropriate arguments.

In order to obtain an exact solution, we suppose that
\begin{equation}\label{40}
W=\lambda \frac{\varphi}{\chi},
\end{equation}
where $\lambda$ is a constant. Hence, we have the following equation for the potential:
\begin{equation}\label{41}
V(\chi,\varphi)=\frac{\lambda}{2 \chi^2}\,\sqrt{\lambda^2(3 \varphi^2-\epsilon)^2-4\varphi^2}.
\end{equation}
From Eqs. (\ref{39}) and (\ref{40}), we can find the following equations:
\begin{equation}\label{42}
\dot \chi =\frac{2 \varphi}{\lambda(3\varphi^2-\epsilon)},\,\,\dot \varphi=-\epsilon \frac{\lambda}{\chi}\,.
\end{equation}
This set of equations can be written down as follows:
\begin{equation}\label{43}
\dot \varphi = B \Big(3 \varphi^2-\epsilon\Big)^{\displaystyle 1/3\epsilon \lambda^2},\,\,\, \chi=-\epsilon\frac{\lambda}{\dot \varphi}\,,
\end{equation}
where $B\ne0$ is a constant of integration. The first equation in  (\ref{45}) is exactly differentiable for several particular values of $\lambda$. Let us consider two simple examples of exact solution for $\epsilon=\pm 1$.

\subsubsection{The case $\epsilon=+1,\,\lambda = 1/\sqrt{3}$}

Integrating the first equation in (\ref{43}) with $\chi>0$, we can obtain that
\begin{equation}\label{44}
\varphi=\frac{1}{\sqrt{3}}\tanh \Big[\sqrt{3}B(t_f-t)\Big],\,\chi=\frac{1}{\sqrt{3}\,B}\cosh^2\Big[\sqrt{3}B(t_f-t)\Big].
\end{equation}
Due to Eqs. (\ref{40}) and (\ref{44}), we have
\begin{equation}\label{45}
H(t)=\frac{B}{\sqrt{3}}\frac{\sinh[\sqrt{3}B(t_f-t)]}{\cosh^3[\sqrt{3}B(t_f-t)]},
\end{equation}
\begin{equation}\label{46}
a(t)=a_0 \exp\left\{\frac{1}{6\cosh^2[\sqrt{3}B(t_f-t)]}\right\}.
\end{equation}
Finally, we can obtain the following expressions for the EoS and deceleration parameters:
\begin{equation}\label{47}
\Big\{\begin{array}{c}
w\\q\end{array}\Big\}=-1-\Big\{\begin{array}{c}
2\\ 3\end{array}\Big\}\times\cosh^2[\sqrt{3}B(t_f-t)]\Big\{\coth^2[\sqrt{3}B(t_f-t)]-3\Big\}.
\end{equation}

\begin{figure}[t]
\centering
\includegraphics[width=90mm,height=7cm]{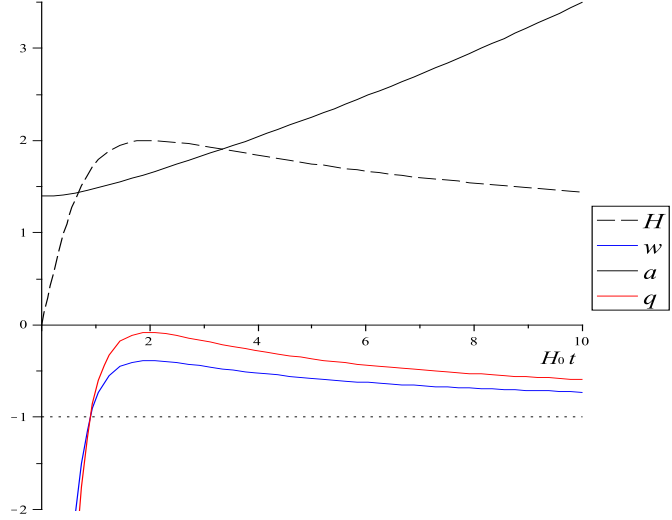}\\
\caption{The plot of the scale factor $a$, the Hubble parameter $H$, the EoS parameter $w$, and the deceleration parameter $q$ versus $H_0 t$ for $\epsilon=-1$, $\lambda=1/\sqrt{3}$, and $H_0=18$.}
\label{pic2}
\end{figure}
\subsubsection{The case $\epsilon = -1,  \lambda = 1/\sqrt{3}$}

In this case, we can integrate equation (\ref{43}) to obtain
\begin{equation}\label{48}
\sqrt{3}\varphi = \Big(H_0 t+\sqrt{1+H_0^2 t^2}\Big)^{1/3}-\Big(H_0 t+\sqrt{1+H_0^2 t^2}\Big)^{-1/3},
\end{equation}
where, for the sake of simplicity, we let the integration constant be zero, and put $B=2H_0/3\sqrt{3}$ with $H_0$ being a constant. Then it is easy to find that $\chi=1/\sqrt{3}\dot \varphi$  and
\begin{equation}\label{49}
H(t)=\frac{H_0}{9\sqrt{1+H_0^2 t^2}}\Big\{\Big(H_0 t+\sqrt{1+H_0^2 t^2}\Big)^{2/3}-\Big(H_0 t+\sqrt{1+H_0^2 t^2}\Big)^{-2/3}\Big\}.
\end{equation}
By using (\ref{21}) and (\ref{49}), we are able to find the EoS  and deceleration parameters as follows:
\begin{equation}\label{50}
\Big\{\begin{array}{c}
w\\q\end{array}\Big\}=-1+\Big\{\begin{array}{c}
2/3\\ 1\end{array}\Big\}\times\left\{\frac{9 H_0 t}{\sqrt{1+H_0^2 t^2}\,g^{(-)}(t)}-\frac{6 g^{(+)}(t)}{[g^{(-)}(t)]^2}\right\},
\end{equation}
where
\begin{equation}\label{51}g^{(\pm)}(t)=\Big(H_0 t+\sqrt{1+H_0^2 t^2}\,\Big)^{2/3}\pm\Big(H_0 t+\sqrt{1+H_0^2 t^2}\,\Big)^{-2/3},
\end{equation}
or
$$
g^{(\pm)}(t)=2\Big\{\begin{array}{c}
\cosh\\\sinh\end{array}\Big\}\Big(\frac{2}{3}\sinh^{-1}H_0t\Big).
$$
By using the obvious equality
$$
\frac{d}{dt}g^{(\pm)}(t)=\frac{2H_0}{3\sqrt{1+H_0^2 t^2}}g^{(\mp)}(t),
$$
it is easy to integrate Eq. (\ref{49}) and obtain the following result for the scale factor:
\begin{equation}\label{52}
a(t)= a_0 \, \exp\Big[\frac{1}{6}g^{(+)}(t)\Big],
\end{equation}
where $a_0$ is a constant of integration. The main features of this solution is shown in Fig. 2. It is interesting to note that, passing through a maximum, then $H\to 0$, and $w,q\to -1$ as $t\to \infty$.

\section{Conclusion}
In this paper, we have studied the flat FRW cosmological models  with interacting quintessence (phantom) and tachyon scalar fields considered as the origin of gravity. We have described theoretical models based on the assumption that the interaction of tachyon field could be described with the help of potential. We have obtained the different types of exact solution for the Friedmann-Robertson-Walker model filled with the interacting  tachyon and quintessence (phantom) scalar fields by employing the so-called first order formalism  procedure.

We have left aside the question of the possibility to reconstruct the potential in the form of an explicit function of the original tachyon field $\phi(t)$. This issue will be studied further.

Of course, all given examples are not the limit of the capacities of the method considered.
We hope that the derived model is the next step in the development of tachyon cosmology, and it can be utilized to describe the dynamics of the evolution of the actual Universe.

\end{document}